\documentclass[letterpaper, 10 pt, conference]{ieeeconf}  

\IEEEoverridecommandlockouts                              

\usepackage[utf8]{inputenc}
\usepackage[T1]{fontenc}
\usepackage{graphicx}
\usepackage{amsmath}
\usepackage{amssymb}
\usepackage{bm}
\usepackage{booktabs}
\usepackage{makecell}
\usepackage{multirow}
\usepackage{subcaption}
\usepackage[hidelinks]{hyperref}
\usepackage{eso-pic}

\title{\LARGE \bf Improving Human-Robot Teamwork in Urban Search and Rescue Through
Episodic Memory of Prior Collaboration}

\author{Taewoon Kim$^{1,2}$, Emma van Zoelen$^{3}$, and Mark Neerincx$^{3}$%
\thanks{$^{1}$HumemAI, The Netherlands. {\tt\small taewoon@humem.ai}}%
\thanks{$^{2}$Vrije Universiteit Amsterdam, The Netherlands}%
\thanks{$^{3}$TNO, The Netherlands. {\tt\small emma.vanzoelen@tno.nl}, {\tt\small mark.neerincx@tno.nl}}%
}

\AddToShipoutPictureFG*{%
  \ifnum\value{page}=1
    \put(\LenToUnit{0.75in},\LenToUnit{0.3in}){\parbox[b]{\dimexpr\paperwidth-1.5in\relax}{\centering\fontsize{6.5}{7.5}\selectfont Accepted for publication at the 2026 IEEE 35th International Conference on Robot and Human Interactive Communication (RO-MAN). DOI: \url{https://doi.org/10.1109/RO-MAN72286.2026.11714607}.\\
    \copyright{} 2026 IEEE. Personal use of this material is permitted. Permission from IEEE must be obtained for all other uses, in any current or future media, including reprinting/republishing this material for advertising or promotional purposes, creating new collective works, for resale or redistribution to servers or lists, or reuse of any copyrighted component of this work in other works.}}%
  \fi
}

\begin{document}

\maketitle
\thispagestyle{empty}
\pagestyle{empty}

\begin{abstract}
Effective human-robot teamwork requires robots to adapt to partners, situations, and
task dynamics from the start of an interaction. In the MATRX Urban Search and Rescue
(USAR) environment, people can externalize collaboration patterns (CPs) they discover
during teamwork through a chat and reflection interface. We study whether a robot can
use such prior team experience to become a better teammate in future interactions. To
this end, we represent historical CPs as knowledge-graph episodic memories and use graph
representation learning with a node-classification objective to identify a
representative and effective memory for reuse. We then initialize the robot with this
memory before a new collaboration episode begins. Across 20 participants and 160
round-level observations, initializing the robot with a single automatically selected
prior CP increases rescue success from 25.7\% to 41.3\% and reduces average task time
by 283 seconds. The strongest gains appear at the
beginning of interaction, suggesting that reusable episodic memory can help robots enter
collaboration with more effective task knowledge and support smoother early teamwork.
\end{abstract}

\section{Introduction}
\label{sec:introduction}

Industrial robots usually work in static settings with limited human
contact~\cite{IFR2023}, but labour shortages and hazardous tasks are driving a shift
toward teammate-style autonomy~\cite{MarketsandMarkets2024}. In these settings, robots
are expected not only to execute actions correctly, but also to coordinate with humans
as partners who have evolving goals, preferences, and working
styles~\cite{Nikolaidis,Smith2024HumanAutonomyTeaming,Haripriyan2024TeamDynamics}.
Disaster response is a particularly demanding example: assistive robots are being
developed to help rescue teams navigate debris, locate survivors, and deliver supplies
in hazardous environments~\cite{Moosavi2024,Betta2024RescuerPerceptions}.

Despite their potential, assistive robots must still interpret human intentions,
adjust to changing situations, and learn from past interactions. Prior work in
human-robot teaming has emphasized mutual adaptation, shared understanding, and
team-level coordination mechanisms that help robots align with their partners over
time~\cite{Nikolaidis,ChaconQuesada2024SharedMentalModels,Haripriyan2024TeamDynamics}.

One approach to studying these issues is the MATRX Urban Search and Rescue (USAR)
environment~\cite{matrx_2023,vanZoelen2021}, a virtual simulation where a human and a
collaborative robot work together to rescue victims. Previous studies have shown that
robots in this setting can adapt to human behavior over time, improving teamwork
efficiency. Notably, when human participants can specify collaboration patterns (CPs)
and communicate through a chat interface, coordination improves and rescue operations
become more effective~\cite{10.1145/3434074.3446354,vanZoelen2021,vanZoelen2022}. In
this context, CPs are defined as human-robot actions conditioned on situations.

Shared memory has long been recognized as a key element in fostering mutual
understanding between humans and robots~\cite{Miraglia2024Shared}. Prior HRI work has
also studied transfer of prior interaction experience for trust-related action
selection and has shown that episodic memory can have especially strong effects early
in interaction~\cite{DiabDemiris2024,VinanziCangelosiGoerick2021}. Building on this
broader line of work, we focus more narrowly on MATRX USAR collaboration patterns
(CPs): we treat previously observed CPs as explicit episodic long-term memories of
prior collaboration and ask whether a robot can carry forward a useful team experience
into a new interaction. To do so, we represent past CPs as knowledge-graph memories and
use graph representation learning with node-type supervision to identify a
representative memory for reuse. Instead of starting each episode with an empty memory
state, the robot begins with a selected prior collaboration pattern. The contribution is
therefore not a new general theory of memory transfer, but a concrete mechanism
for initializing a robot with an inspectable prior team experience in this particular
USAR setting. Because the reused experience remains explicit rather than opaque policy
parameters, later robot behavior can still be inspected and revised.

To summarize, our main contributions are as follows:

\begin{itemize}
    \item \textbf{Episodic Team Memory Representation:} We represent collaboration
    patterns (CPs) from prior human-robot interactions as knowledge-graph episodic
    memories, preserving situational context, action structure, and task outcomes in a
    form that can be reused across episodes.
    
    \item \textbf{Selecting Reusable Prior Experience:} We use graph representation
    learning with node-type supervision to organize historical CPs and identify a
    representative prior collaboration pattern to initialize the robot before a new
    interaction begins.

    \item \textbf{Improved Early-Team Performance:} Initializing the robot with a
    selected prior memory raises rescue success from 25.7\% to 41.3\% (a 60.7\% relative
    improvement), reduces average task time by 283 seconds, and yields the largest gains
    at the beginning of collaboration, when teams are still orienting to the task.
        
\end{itemize}

\section{Background}
\label{sec:background}

\subsection{MATRX Urban Search and Rescue (USAR)}
\label{sec:usar}

\begin{figure}[tb]
\centering
\includegraphics[width=\columnwidth]{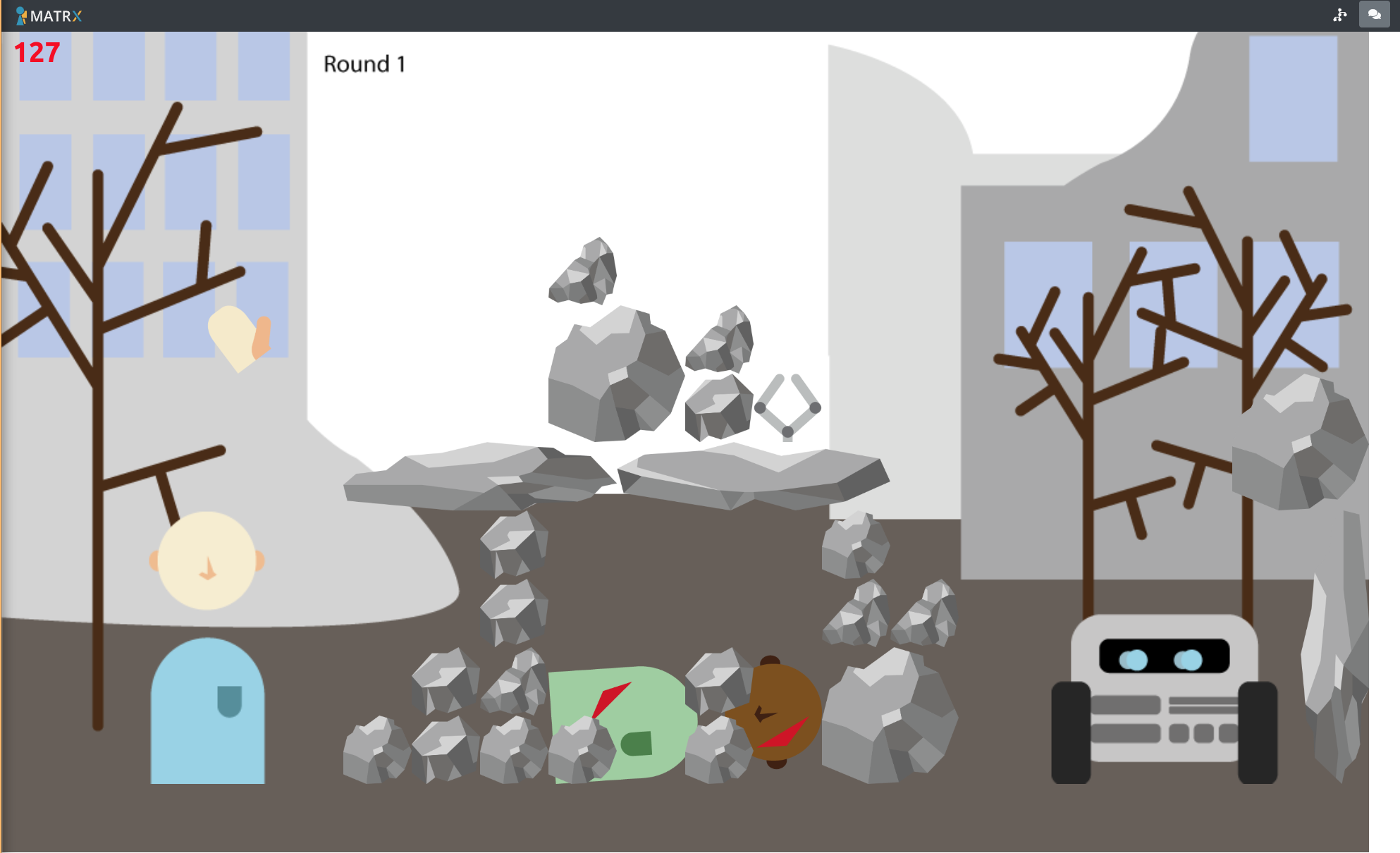}
\caption{A screenshot of the MATRX USAR simulation. The human user is shown on the left, while the robot is on the right. In the upper right corner, users can click icons to define a collaboration pattern (CP) or open the chat interface.}
\label{fig:usar-main}
\end{figure}

\begin{figure}[tb]
\centering

\begin{subfigure}[b]{\columnwidth}
    \centering
    \includegraphics[width=\columnwidth]{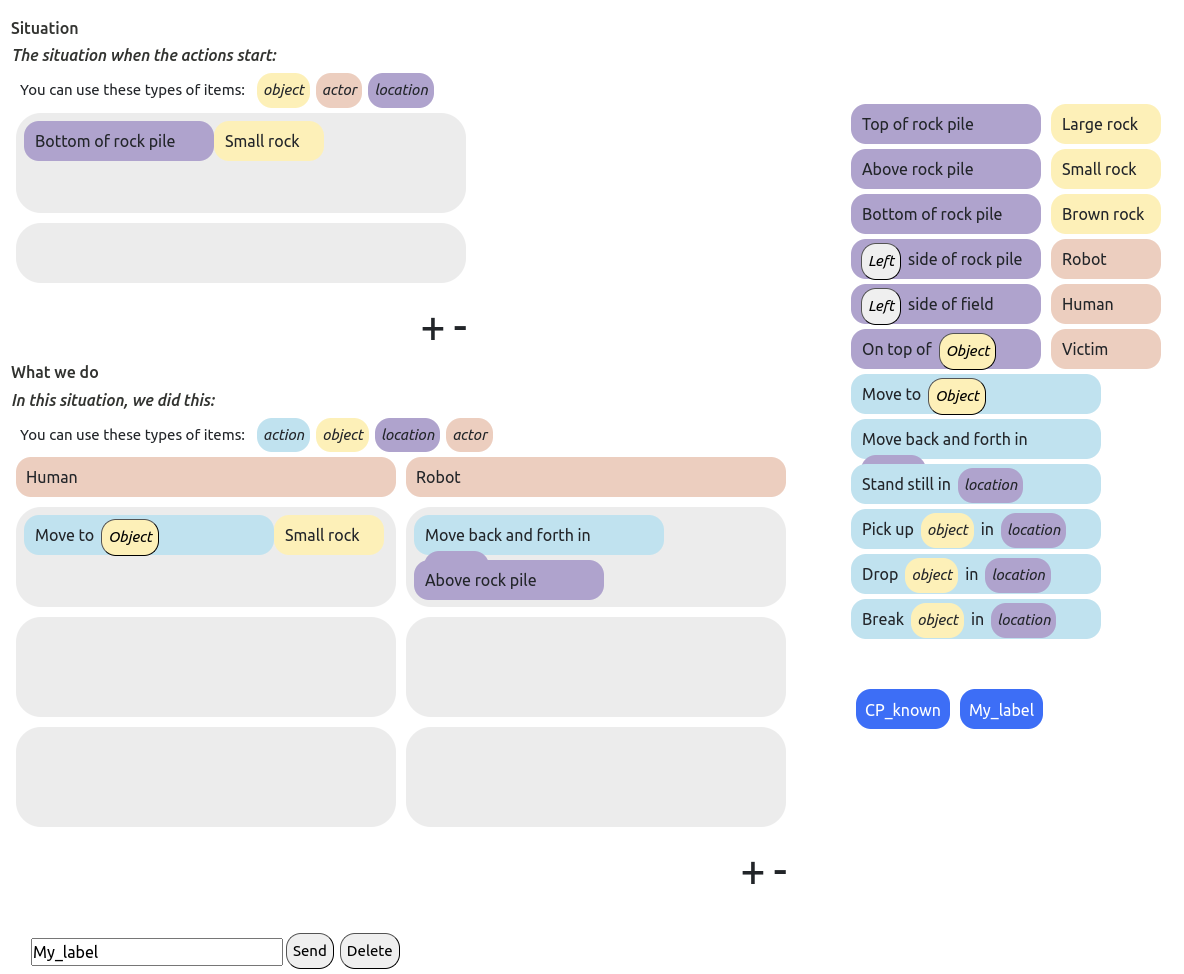}
    \caption{Users can drag and drop boxes (representing objects, actors, and locations) to document the emergent CPs they observe with the robot.}
\end{subfigure}

\vspace{1em}

\begin{subfigure}[b]{\columnwidth}
    \begin{minipage}[b]{0.49\columnwidth}
        \centering
        \includegraphics[width=\linewidth]{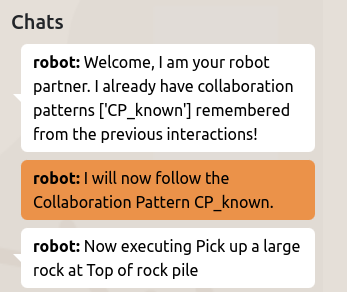}
    \end{minipage}
    \hfill
    \begin{minipage}[b]{0.49\columnwidth}
        \small
        This chat interface enables the robot to communicate updates
        and inform human users about ongoing events.
    \end{minipage}

    \caption{The chat interface.}
\end{subfigure}

\caption{User interaction features in the MATRX USAR environment.
(a) The CP interface. (b) The chat interface.}
\label{fig:usar-interaction}
\end{figure}

The MATRX Urban Search and Rescue (USAR) environment~\cite{matrx_2023,vanZoelen2021}
simulates a USAR scenario where a collaborative virtual robot and a human must work
together to rescue a victim. Figure~\ref{fig:usar-main} provides an example screenshot
of the environment, while Figure~\ref{fig:usar-interaction} illustrates the user
interface used for human-robot communication. The CPs were originally stored in
TypeDB~\cite{typeql2024}.

The team objective is to free a buried victim by removing rubble in front of the victim
and opening an access route from one side, while minimizing elapsed time and avoiding
additional harm from falling rocks~\cite{vanZoelen2021}. The task is intentionally
asymmetric: the robot can manipulate both small and large rocks, whereas the human is
more limited physically but can better anticipate side effects such as rock falls and
unsafe removals~\cite{vanZoelen2021,vanZoelen2025thesis}. This asymmetry makes the task
genuinely collaborative rather than a simple division of identical actions.

The study includes one practice round followed by eight real rounds: five easier rounds
and three harder rounds introducing a brown rock whose placement can make rescue
impossible if handled poorly~\cite{vanZoelen2021}. This repeated structure lets human
and robot adapt over time so that emergent strategies and collaboration patterns can be
observed across rounds~\cite{vanZoelen2021,vanZoelen2025thesis}. Participants can define
CPs that remain available unless they are later modified or removed. Later work added
reflective communication support so that teams could explicitly externalize and discuss
such patterns through a shared ontology and chat interface~\cite{10.1145/3434074.3446354,vanZoelen2022}.

\subsection{Knowledge Graphs (KGs)}
\label{sec:kg}

A KG is a directed graph $\mathcal{G} = (\mathcal{V}, \mathcal{R}, \mathcal{E})$, where
$\mathcal{V}$ represents entities, $\mathcal{R}$ denotes relations, and $\mathcal{E}
\subseteq \mathcal{V} \times \mathcal{R} \times \mathcal{V}$ consists of edges that
define the relationships between entities. Each edge $(v_i, r_k, v_j)$ indicates that a
relation $r_k$ exists from entity $v_i$ to entity $v_j$~\cite{10.1145/3447772}. While
RDF graphs emphasize subject-predicate-object triples and
interoperability~\cite{Lanthaler:14:RCA}, we use a property graph model implemented in
JanusGraph~\cite{janusgraph_2024}, which allows both nodes and edges to carry attributes
and supports efficient querying with Gremlin, making it well-suited for capturing the
situational and action structure of CPs.

\subsection{Graph Neural Networks (GNNs)}
\label{sec:gnn}

GNNs are deep learning models designed for graph-structured data~\cite{4700287}. They
update each node's representation by aggregating features from its neighbors through
successive message-passing layers, making them well-suited for tasks involving complex
relational structures such as KG reasoning and graph representation learning. In our
setting, this means that a node representation can reflect not only local attributes but
also the surrounding situation and action context through neighborhood-based graph
convolution. In the original Graph Convolutional Network (GCN), neighbor information is
aggregated without distinguishing edge or relation types~\cite{DBLP:journals/corr/KipfW16}.
Relational Graph Convolutional Networks (RGCNs) extend this idea by introducing
relation-specific transformations for typed edges~\cite{schlichtkrull2017modelingrelationaldatagraph}.
For an RGCN layer, a standard update is
\[
\mathbf{h}_v^{(\ell+1)} = \sigma\!\left( \sum_{r \in \mathcal{R}} \sum_{u \in \mathcal{N}_v^r}
\frac{1}{c_{v,r}} \mathbf{W}_r^{(\ell)} \mathbf{h}_u^{(\ell)} + \mathbf{W}_0^{(\ell)}
\mathbf{h}_v^{(\ell)} \right),
\]
where $\mathbf{h}_v^{(\ell)}$ is the representation of node $v$ at layer $\ell$,
$\mathcal{N}_v^r$ denotes the neighbors connected to $v$ by relation type $r$,
$\mathbf{W}_r^{(\ell)}$ is a relation-specific weight matrix, $c_{v,r}$ is a
normalization constant, $\mathbf{W}_0^{(\ell)}$ is the weight matrix for the self-node
contribution, and $\sigma$ is a nonlinearity. In our setting, the edge labels serve as
relation types that determine how messages are passed; we do not learn separate
edge-state updates beyond these relation-specific transformations.

\section{Methodology}
\label{sec:methodology}

\subsection{Ontology Engineering}
\label{sec:ontology}

Because CP graphs have variable topology, we represent them as property graphs rather
than force‑fitting them into fixed‑length Euclidean vectors, thus retaining full
relational context.

Our ontology defines six entity types: \texttt{robot}, \texttt{participant},
\texttt{cp}, \texttt{situation}, \texttt{robot\_action}, and \texttt{human\_action}.
Each entity is represented as a node with associated properties. Relationships between
entities are captured through four edge families: \texttt{has\_cp},
\texttt{has\_situation}, \texttt{has\_robot\_action}, and \texttt{has\_human\_action}.
For descriptive summaries, we annotate action edges with their sequence index (e.g.,
\texttt{has\_robot\_action\_0}) to show their order within a CP, but these indices do
not change the underlying ontology.

For graph learning, we make a more fine-grained distinction and treat these indexed
action relations as different relation labels. The reason is simple: the encoder should
be able to tell apart an early action from a later one, since action order is part of
the collaboration pattern. If all robot-action edges shared one label, then ``first
robot action'' and ``later robot action'' would look identical to the RGCN. Therefore,
the encoder uses nine directed relation labels in total: one \texttt{has\_situation}
label, three \texttt{has\_human\_action\_k} labels, and five
\texttt{has\_robot\_action\_k} labels.

\begin{figure}[tb]
\centering
\begin{subfigure}[b]{\columnwidth}
    \centering
    \includegraphics[width=0.7\columnwidth]{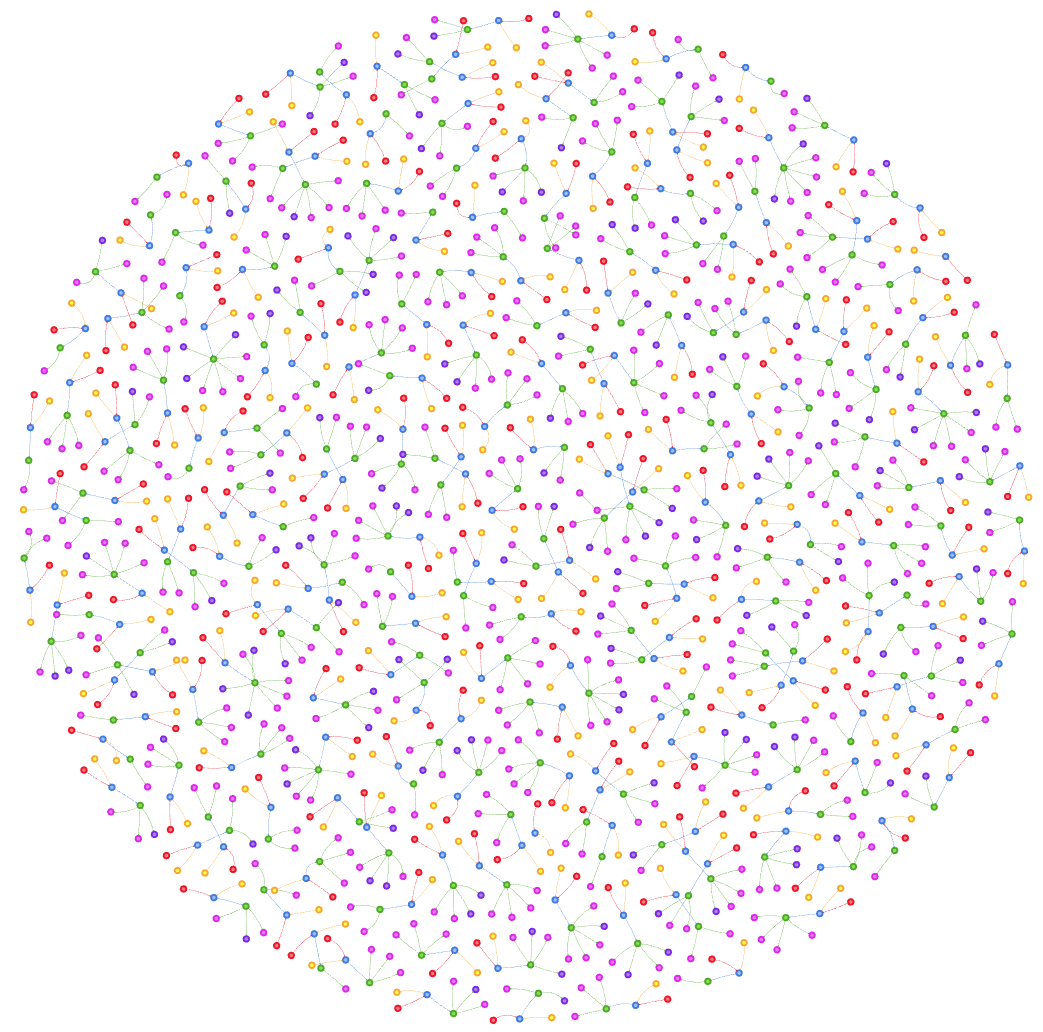}
    \caption{A visualization of collected 209 collaboration patterns (CPs) as 209 knowledge graphs (KGs).}
\end{subfigure}

\vspace{1em}

\begin{subfigure}[b]{\columnwidth}
    \centering
    \includegraphics[width=0.75\columnwidth]{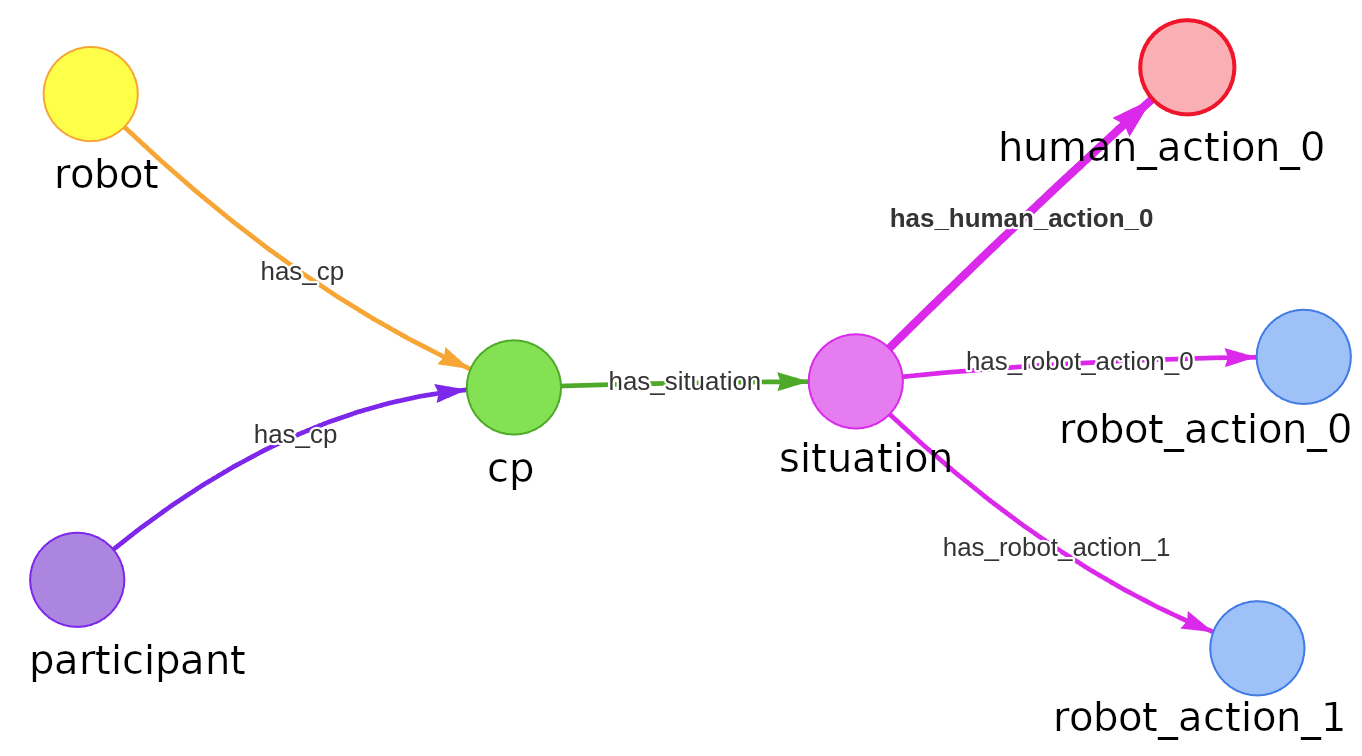}
    \caption{A visualization of an example CP. This knowledge graph (KG) has seven nodes: one \texttt{robot}, \texttt{participant}, \texttt{cp}, and \texttt{situation}, along with one \texttt{human\_action} and two \texttt{robot\_action} nodes. While all CPs share the same core structure, the number of \texttt{human\_action} and \texttt{robot\_action} nodes varies, as annotators define action sequences for the human and robot. This graph representation naturally accommodates such variable-length collaborative routines.}
    \label{fig:cp_36_visualized}
\end{subfigure}

\caption{Visualizations of collaboration patterns (CPs) represented as knowledge graphs (KGs). (a) The entire set of 209 CPs. (b) An example CP illustrating its structural components.}
\label{fig:cp_visualization}
\end{figure}

From the previous studies, we collected 209 CPs, visualized in
Figure~\ref{fig:cp_visualization} using the JanusGraph-Visualizer~\cite{janusgraph_2024}.

Table~\ref{tab:cp_properties} lists the node properties of the example CP in
Figure~\ref{fig:cp_36_visualized}. The \texttt{cp} node properties ``time\_elapsed'',
``remaining\_rocks'', ``victim\_harm'', and ``success'' serve as round-level
performance metrics.

\begin{table}[tb]
    \centering
    \renewcommand{\arraystretch}{1.2}
    \begin{tabular}{p{2.5cm} p{5.5cm}}
        \toprule
        \textbf{Type} & \textbf{Properties} \\
        \midrule
        \texttt{robot} &  \\
        \midrule
        \texttt{participant} & participant\_number: 4,081 \\
        \midrule
        \texttt{cp} & time\_elapsed: 3,000, \newline event\_time: ``2024-05-07T11:26:11'', \newline cp\_num: 36, \newline ticks\_lasted: 1,917, \newline participant\_num: 4,081, \newline remaining\_rocks: 32, \newline cp\_name: ``top brown'', \newline victim\_harm: 0, \newline success: false \newline round\_num: 6 \\
        \midrule
        \texttt{situation} & location: ``Top of rock pile'', \newline object: ``Brown rock'' \\
        \midrule
        \texttt{human\_action\_0} & location: ``Above rock pile'', \newline action: ``Stand still in \textless location\textgreater'' \\
        \midrule
        \texttt{robot\_action\_0} & location: ``Top of rock pile'', \newline object: ``Brown rock'', \newline action: ``Pick up \textless object\textgreater in \textless location\textgreater'' \\
        \midrule
        \texttt{robot\_action\_1} & location: ``\textless Right\textgreater side of field'', \newline object: ``Brown rock'', \newline action: ``Drop \textless object\textgreater in \textless location\textgreater'' \\
        \bottomrule
    \end{tabular}
    \caption{The node properties of the example CP.}
    \label{tab:cp_properties}
\end{table}

\subsection{Graph Representation Learning on CP Graphs}
\label{sec:graph_learning}

Choosing a CP by raw score alone ignores the situational and action context that
produced that score. A high-performing CP in one scenario may not generalize to others.
To address this, we cluster CPs using graph representation learning, capturing both
performance and structural similarities. This enables us to identify a representative
and effective CP for reuse. 

Since our CPs are structured as KGs, we use a Relational Graph Convolutional Network
(RGCN)~\cite{schlichtkrull2017modelingrelationaldatagraph} to learn meaningful
representations. Unlike standard GNNs, RGCN incorporates edge types (e.g.,
\texttt{has\_cp}, \texttt{has\_human\_action}), which capture essential relational
information. We train the encoder with a node-classification objective using a single
linear classifier on top of the final RGCN layer. The ten node classes correspond to
one \texttt{cp} class, one \texttt{situation} class, three human-action-stage classes,
and five robot-action-stage classes. Our representation learning therefore uses
supervision derived from the CP schema rather than a fully self-supervised objective.
In the implementation, a ReLU activation is applied between the two RGCN layers, while
the input projection and the final classification head are linear.

As shown in Table~\ref{tab:cp_properties}, node properties contain various features and
must be transformed into fixed-length vectors. For
\texttt{cp} nodes, we concatenate six numerical features (``time\_elapsed'',
``ticks\_lasted'', ``remaining\_rocks'', ``victim\_harm'', ``success'', and
``round\_num''), normalizing each by its maximum value to scale them between 0 and 1.
For \texttt{situation}, \texttt{human\_action}, and \texttt{robot\_action} nodes, we
concatenate string attributes into natural language sentences (e.g.,
\texttt{human\_action\_0} in the example would be ``Action: Stand still in \textless
location\textgreater. Location: Above rock pile.'') and encode them using Sentence
Bidirectional Encoder Representations from Transformers
(Sentence-BERT)~\cite{reimers2019sentencebertsentenceembeddingsusing} to obtain
numerical vector representations. Since the \texttt{cp} node vectors are shorter than
the others, we apply a learnable projection matrix to transform them to the same
dimensionality as the vectors for \texttt{situation}, \texttt{human\_action}, and
\texttt{robot\_action} nodes. Concretely, if node $v$ is a low-dimensional
\texttt{cp} node with input $\mathbf{x}_v \in \mathbb{R}^{d_s}$, we first project it as
$\tilde{\mathbf{x}}_v = \mathbf{W}_{\mathrm{up}} \mathbf{x}_v$,
where $\mathbf{W}_{\mathrm{up}} \in \mathbb{R}^{d_b \times d_s}$ maps it into the common
$d_b$-dimensional feature space used by the other node types.

We optimize RGCN using cross-entropy loss for node classification. Let \( f_{\theta} \)
represent the entire model, including RGCN and the linear classifier. Given a CP (KG) \(
\mathcal{G} \) with node-feature matrix \( X \) and graph connectivity \( A \), the predicted
class distribution for a node \( v \) is computed from the final RGCN embedding
\(\mathbf{h}_v^{(2)}\) as $\mathbf{z}_v = \mathbf{W}_{\mathrm{cls}} \mathbf{h}_v^{(2)}$
and $\hat{\mathbf{y}}_v = \mathrm{softmax}(\mathbf{z}_v)$, where
$\mathbf{W}_{\mathrm{cls}}$ denotes the single linear classification head.
The loss function is:

\[
\mathcal{L} = - \mathbb{E}_{(v, c) \sim \mathcal{D}_\mathrm{train}} \left[ \log \hat{y}_{v, c} \right]
\]

where \(\mathcal{D}_\mathrm{train}\) represents the empirical distribution of training
samples, \(\hat{y}_{v,c}\) denotes the predicted probability assigned to the true class
\(c\) of node \(v\), and \(C\) is the number of node classes.

\begin{figure}[tb]
\centering
\includegraphics[width=\columnwidth]{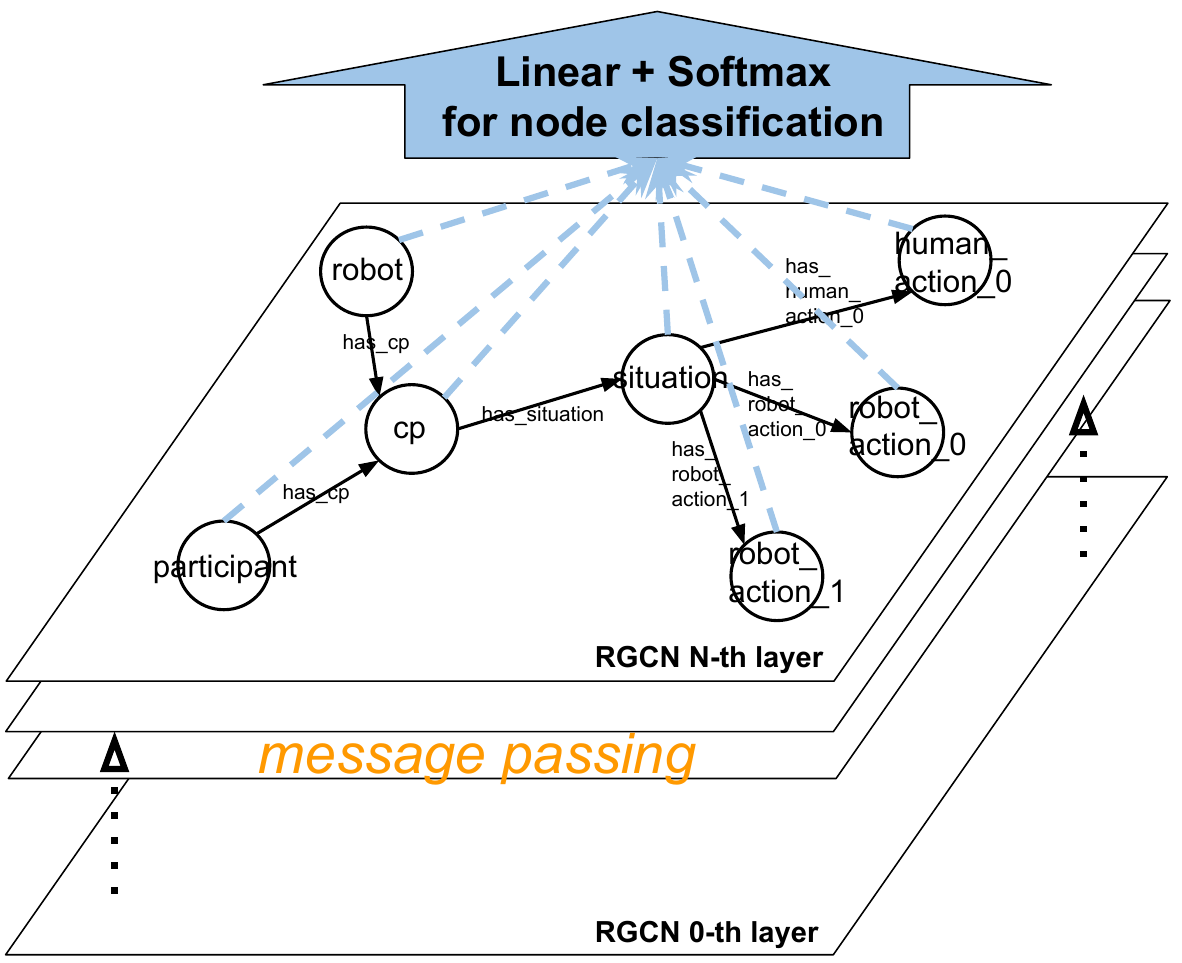}
\caption{A forward pass of the RGCN-based neural network. Low-dimensional \texttt{cp} node features are first linearly projected to the common feature space, and a single linear classifier with softmax is applied to the final RGCN node embeddings to compute the cross-entropy loss.}
\label{fig:rgcn}
\end{figure}

Action stages are treated as distinct node classes so that the encoder can distinguish
early from later steps within a collaboration pattern. In total, we use \(C = 10\)
node classes. See
Figure~\ref{fig:rgcn} for a visualization of one forward pass of the RGCN-based neural
network. After training, we obtain a graph-level embedding for each CP graph
\(\mathcal{G}_i = (\mathcal{V}_i, \mathcal{E}_i)\) by mean pooling its final node
embeddings,
\[
\mathbf{g}_i = \frac{1}{|\mathcal{V}_i|} \sum_{v \in \mathcal{V}_i} \mathbf{h}_v^{(2)},
\]
where \(|\mathcal{V}_i|\) is the number of nodes in graph \(\mathcal{G}_i\). These
pooled graph embeddings are the vectors that we cluster with K-means.

After this representation-learning phase, we use the learned graph representations to
cluster the CPs. From these clusters, we select one centroid-nearest representative CP,
which we then use to initialize the robot's episodic memory. This selected CP is not a
learned control policy. Rather, it is a structured memory item containing a situation
and an associated action sequence that is preloaded before a new trial begins. During
execution, the robot continues to use the existing MATRX control framework, but it can
reuse the preloaded CP when the current situation matches the stored condition. We then
conduct human-robot interaction experiments in the same MATRX collaborative environment
to evaluate the impact of this memory augmentation on task performance metrics. The
transferred prior remains a human-readable situation-action structure that can be
inspected, discussed, and revised rather than a latent policy update. Because the
encoder is trained for node-type classification rather than
downstream team performance, we use the resulting embeddings only to group and compare
structurally similar CPs during heuristic selection, not to directly rank memories by
their expected transfer value.

\section{Experiments}
\label{sec:experiments}

\subsection{Training and Clustering}

\begin{figure}[tb]
\centering
\includegraphics[width=\columnwidth]{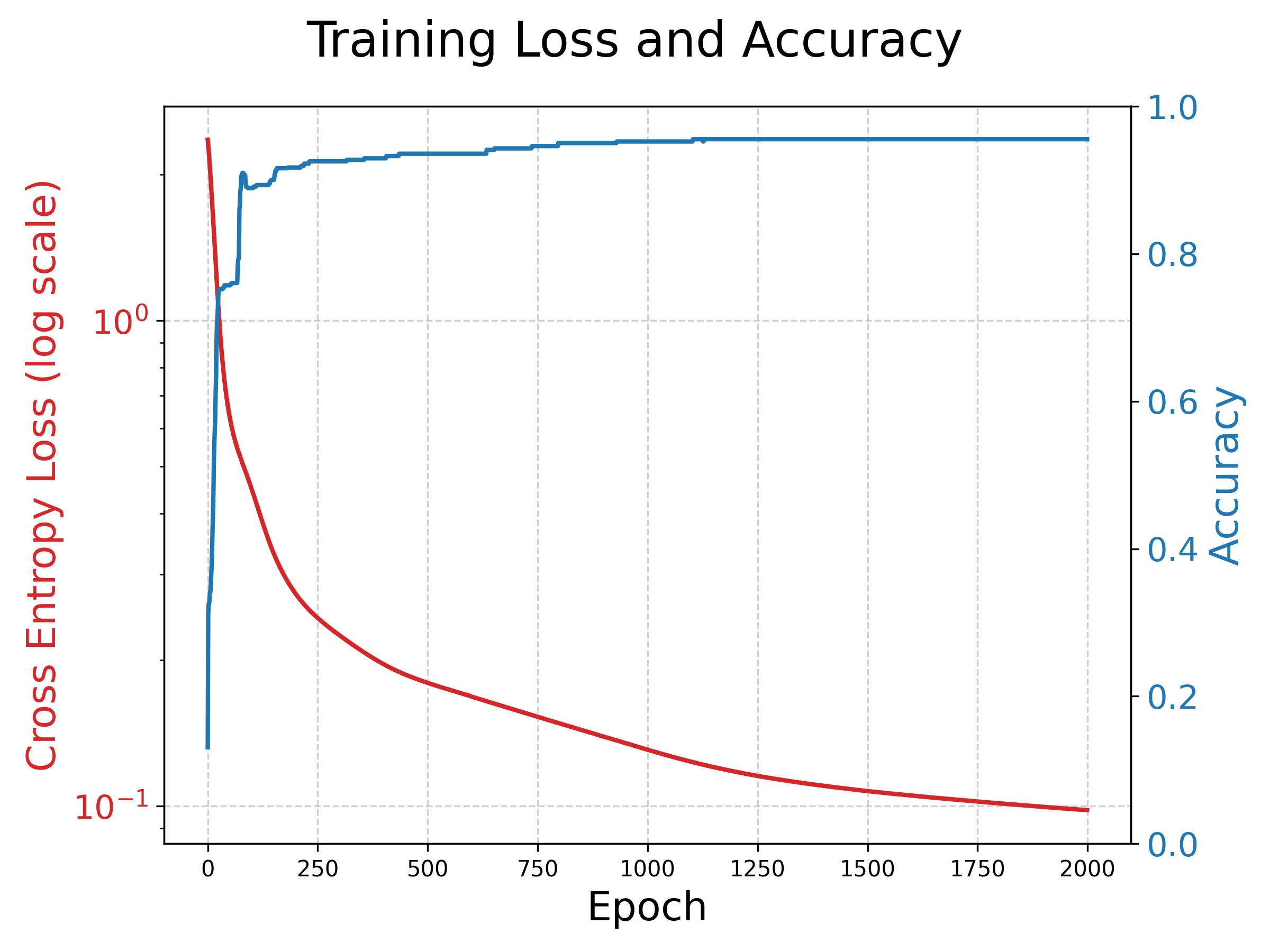}
\caption{Cross entropy loss and accuracy of the training data. After 2,000 epochs, the loss dropped to 0.0948 while the accuracy went up to 95.5\%.}
\label{fig:training}
\end{figure}

The two‑layer RGCN (128,874 parameters, ReLU activations) trained for 2,000 epochs on
all 209 CP graphs in a single 30s CPU run; code, hyper‑parameters, supplementary
material, and raw data are available at \url{https://github.com/humemai/co-learning}. See
Figure~\ref{fig:training} for the training loss, which steadily decreases over epochs,
while the accuracy increases, indicating effective learning and convergence of the
model.

\begin{figure}[tb]
\centering

\begin{subfigure}[b]{\columnwidth}
    \centering
    \includegraphics[width=0.85\columnwidth]{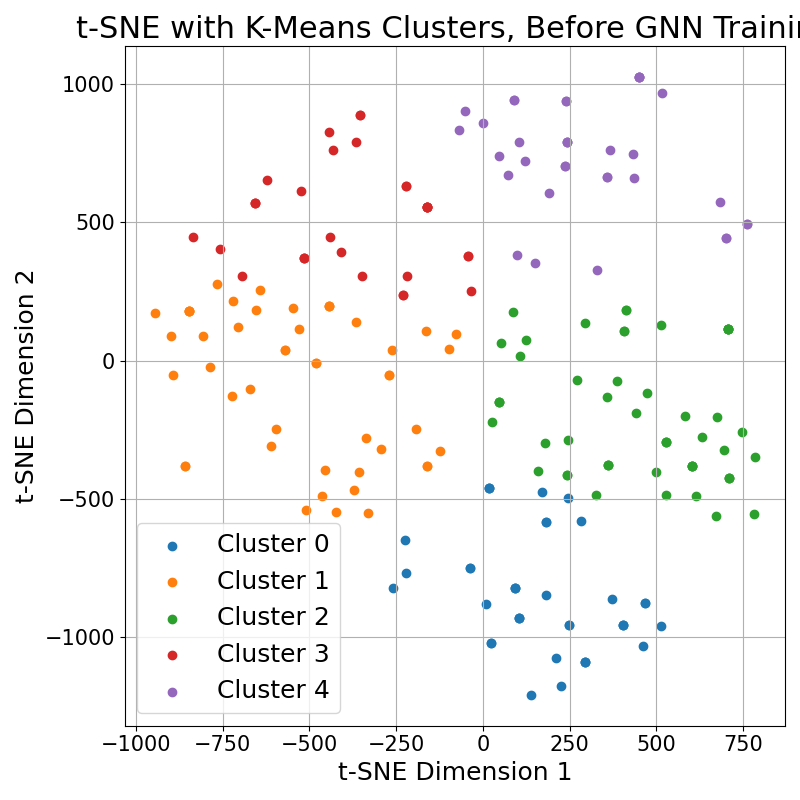}
    \caption{5 clusters of CPs before training. Since there are not yet GNN-based vector representations for the CPs, we used the \texttt{cp} node features as vectors.}
\end{subfigure}

\vspace{1em}

\begin{subfigure}[b]{\columnwidth}
    \centering
    \includegraphics[width=0.85\columnwidth]{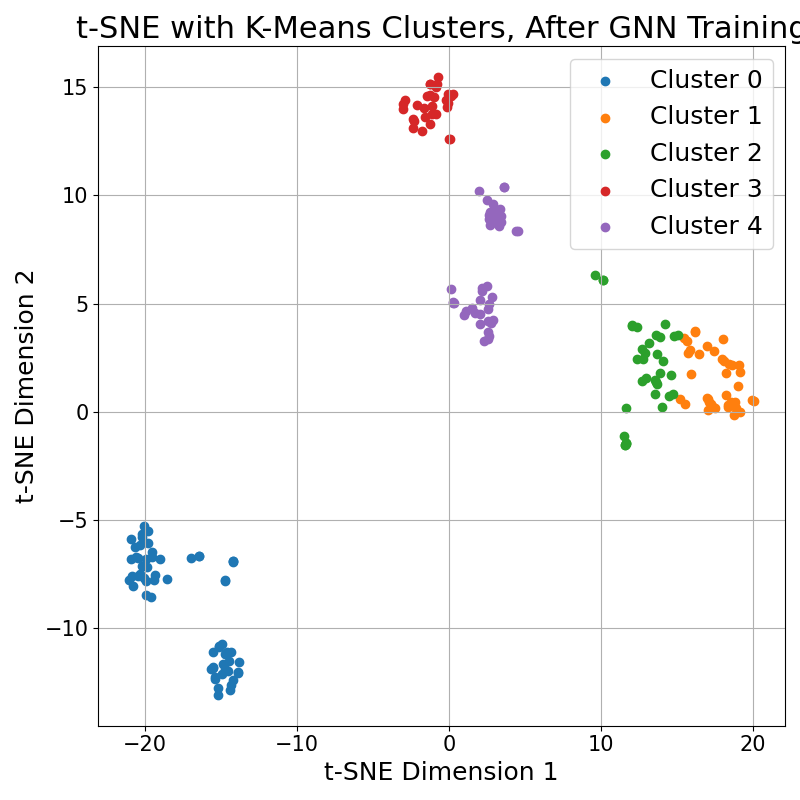}
    \caption{5 clusters of CPs after training. The vectors used here are the means of the last RGCN layer node representations.}
\end{subfigure}

\caption{K-means clustering results before and after training, visualized using t-SNE. $K=5$ was chosen.}
\label{fig:cluster}
\end{figure}

After training, we obtain the final vector representation of each CP by averaging the
node representations (before the classification layer) within each graph. We then apply
the K-means clustering algorithm to group CPs based on Euclidean distance.
Figure~\ref{fig:cluster} illustrates the clusters before and after training, with
t-distributed stochastic neighbor embedding (t-SNE)~\cite{vanDerMaaten2008TSNE} used for
2D visualization. The learned RGCN representations produce clusters that are more
interpretable than those obtained from raw \texttt{cp} node features alone. The number
of CPs in the clusters are 62, 38, 36, 30, and 43, respectively. We set $K=5$, where
$K$ is the number of K-means clusters, based on exploratory inspection of cluster
interpretability and balance; this was a heuristic design choice rather than a tuned
optimum, and we did not optimize $K$ against downstream team performance.

Table~\ref{tab:cluster_stats} presents the performance metrics for each cluster.
Notably, Cluster 1 exhibits a higher success rate, suggesting that participants using
CPs from this cluster were more likely to rescue the victim. While the other metrics
remain important, we emphasize success rate since saving the victim is considered the
most critical outcome.

\begin{table}[tb]
    \centering
    \renewcommand{\arraystretch}{1.2}
    \resizebox{\columnwidth}{!}{ 
    \begin{tabular}{lcccc}
        \toprule
        \textbf{Cluster} & \makecell{\textbf{Time} \\ \textbf{Elapsed}} & \makecell{\textbf{Remaining} \\ \textbf{Rocks}} & \makecell{\textbf{Victim} \\ \textbf{Harm}} & \makecell{\textbf{Success} \\ \textbf{Rate}} \\
        \midrule
        Cluster 0 & 2,716 (+120) & 16 (+1.0) & 291 (-14) & 24\% (-3\%) \\
        Cluster 1 & 2,133 (-462) & 11 (-3.9) & 318 (+11) & 42\% (+15\%) \\
        Cluster 2 & 2,607 (+12) & 11 (-3.2) & 311 (+4) & 36\% (+9\%) \\
        Cluster 3 & 2,869 (+273) & 18 (+3.9) & 246 (-60) & 17\% (-11\%) \\
        Cluster 4 & 2,629 (+33) & 16 (+1.9) & 355 (+49) & 19\% (-9\%) \\
        \bottomrule
    \end{tabular}
    }
    \caption{Cluster statistics (higher is better for success rate and lower is better for the others.), with differences from the mean in parentheses.}
    \label{tab:cluster_stats}
\end{table}

Table~\ref{tab:cluster_edges} presents the edge types and their frequencies for the CPs
in each cluster. Notably, the action edges in Cluster 1 are simpler than those in other
clusters, as it contains fewer distinct action types. Of the 78 total edges across all
CPs in Cluster 1, the majority involve two consecutive robot actions. This suggests that
Cluster 1 not only represents a simpler interaction pattern but also grants greater
autonomy to the robot. While robot actions are typically conditioned on human actions,
the lower presence of human action edges in this cluster indicates that the robot's
behavior is less dependent on human input.

\begin{table}[tb]
    \centering
    \renewcommand{\arraystretch}{1.2}
    \setlength{\tabcolsep}{8pt}
    \begin{tabular}{llc}
        \toprule
        \textbf{Cluster} & \textbf{Edge label} & \textbf{Count (\%)} \\
        \midrule
        \multirow{7}{*}{Cluster 0} 
        & has\_robot\_action\_0 & 62 (55.86\%) \\
        & has\_human\_action\_0 & 30 (27.03\%) \\
        & has\_human\_action\_1 & 8 (7.21\%) \\
        & has\_robot\_action\_2 & 3 (2.70\%) \\
        & has\_robot\_action\_3 & 3 (2.70\%) \\
        & has\_robot\_action\_4 & 3 (2.70\%) \\
        & has\_human\_action\_2 & 2 (1.80\%) \\
        \midrule
        \multirow{3}{*}{Cluster 1} 
        & has\_robot\_action\_1 & 38 (48.72\%) \\
        & has\_robot\_action\_0 & 38 (48.72\%) \\
        & has\_robot\_action\_2 & 2 (2.56\%) \\
        \midrule
        \multirow{6}{*}{Cluster 2} 
        & has\_robot\_action\_1 & 36 (31.30\%) \\
        & has\_robot\_action\_0 & 36 (31.30\%) \\
        & has\_robot\_action\_2 & 31 (26.96\%) \\
        & has\_robot\_action\_3 & 7 (6.09\%) \\
        & has\_human\_action\_1 & 4 (3.48\%) \\
        & has\_human\_action\_2 & 1 (0.87\%) \\
        \midrule
        \multirow{4}{*}{Cluster 3} 
        & has\_human\_action\_0 & 30 (38.46\%) \\
        & has\_robot\_action\_1 & 30 (38.46\%) \\
        & has\_robot\_action\_2 & 16 (20.51\%) \\
        & has\_robot\_action\_3 & 2 (2.56\%) \\
        \midrule
        \multirow{7}{*}{Cluster 4} 
        & has\_human\_action\_0 & 43 (26.22\%) \\
        & has\_robot\_action\_1 & 43 (26.22\%) \\
        & has\_robot\_action\_0 & 42 (25.61\%) \\
        & has\_human\_action\_1 & 20 (12.20\%) \\
        & has\_robot\_action\_2 & 7 (4.27\%) \\
        & has\_robot\_action\_3 & 7 (4.27\%) \\
        & has\_human\_action\_2 & 2 (1.22\%) \\
        \bottomrule
    \end{tabular}
    \caption{Edge label distributions across clusters.}
    \label{tab:cluster_edges}
\end{table}

\subsection{Representative CP Selection}

Considering both success rate and structural simplicity, we heuristically selected
Cluster 1 as the cluster from which to draw a reusable CP. It represents a minimal yet
effective interaction pattern, where the robot operates with greater autonomy while
maintaining efficiency. We then selected the CP closest to the cluster center (by
Euclidean distance) as our final choice, using it as the centroid-nearest representative
of that cluster as a pragmatic exemplar-selection procedure rather than claiming
it to be globally optimal. In other words, the final choice combines learned embeddings
with manual inspection of cluster-level statistics; it is intended to demonstrate one
feasible way to select a reusable CP rather than a fully automatic or provably best
selection rule.

Table~\ref{tab:final_cp} presents the final selected CP. Notably, this CP consists of
only two consecutive robot actions without any human actions. This suggests that, in
this scenario, the robot operates independently, executing its tasks without relying on
human input. Compared to other CPs that typically involve human actions, this CP grants
the robot greater autonomy in decision-making and task execution. Specifically, this CP
directs the robot to move large rocks near the victim. Since only the robot can handle
large rocks, this pattern effectively delegates the heavy lifting to the robot while
allowing the human to focus on other tasks as they see fit.

\begin{table}[tb]
    \centering
    \renewcommand{\arraystretch}{1.2}
    \setlength{\tabcolsep}{5pt}
    \begin{tabular}{lp{5cm}}
        \toprule
        \textbf{Type} & \textbf{Details} \\
        \midrule
        \texttt{situation} & Location: ``Top of rock pile'' \newline
                   Object: ``Large rock'' \\
        \midrule
        \texttt{robot\_action\_0} & Location: ``Top of rock pile'' \newline
                         Object: ``Large rock'' \newline
                         Action: ``Pick up \textless object\textgreater{} in \textless location\textgreater{}'' \\
        \midrule
        \texttt{robot\_action\_1} & Location: ``\textless Right\textgreater{} side of field'' \newline
                         Object: ``Large rock'' \newline
                         Action: ``Drop \textless object\textgreater{} in \textless location\textgreater{}'' \\
        \bottomrule
    \end{tabular}
    \caption{The final selected collaboration pattern (CP).}
    \label{tab:final_cp}
\end{table}

\subsection{Human-Subject Evaluation}

We initialize the robot's memory with the final CP and conduct the MATRX USAR experiment
under the same conditions as the previous study, with the only difference being the
inclusion of the preloaded memory. The historical memory pool contains 209 previously
collected CPs from earlier MATRX studies and is treated here as a fixed offline memory
source. The robot is initialized with only one selected CP at test time. The same
selected CP is reused for all participants in the memory-initialized condition, so the
evaluation tests transfer of one shared prior team experience rather than per-user
personalization. We recruited 20 student participants from our university. All
experiments were conducted online.\footnote{The experiment involving human participants
was approved by the Human Research Ethics Committee of Delft University of Technology.}

\subsection{Results and Failure Cases}

Table~\ref{tab:comparison} shows average improvements for the memory-initialized robot
in success rate, elapsed time, and remaining rocks, but not in every metric or every
round. Victim harm is higher on average in the memory-initialized condition, likely due
to accidental rock falls when removing large rocks. Across the 160 round-level observations, the
success rate increased from 25.7\% to 41.3\% (a 60.7\% relative improvement). Because
each participant contributes multiple rounds, these 160 round-level observations are
not independent. We therefore report this as an observed round-level difference rather
than as a formal participant-level comparison.
The benefit is therefore not uniform across all metrics or all rounds: as noted, victim
harm is higher on average in the memory-initialized condition, and performance drops in
Rounds 7 and 8. This pattern suggests a trade-off in which the transferred CP can improve early
coordination and efficiency while still introducing safety-relevant failure modes when
heavy-debris actions are reused in harder scenarios.
The most significant improvement occurs in Round 1, when participants are still
unfamiliar with the environment. This suggests that the benefit appears before
substantial within-session human adaptation can occur, giving the team a more
coordinated starting point.

\begin{table*}[tb]
    \centering
    \renewcommand{\arraystretch}{1.2}
    \setlength{\tabcolsep}{6pt}
    \resizebox{\textwidth}{!}{ 
    \begin{tabular}{c|cc|cc|cc|cc}
    \toprule
    \multirow{2}{*}{Round} 
    & \multicolumn{2}{c|}{\textbf{Time Elapsed}} 
    & \multicolumn{2}{c|}{\textbf{Remaining Rocks}} 
    & \multicolumn{2}{c|}{\textbf{Victim Harm}} 
    & \multicolumn{2}{c}{\textbf{Success Rate}} \\
    & \textbf{\makecell{W/o Memory \\ Initialized}} & \textbf{\makecell{With Memory \\ Initialized}} 
    & \textbf{\makecell{W/o Memory \\ Initialized}} & \textbf{\makecell{With Memory \\ Initialized}} 
    & \textbf{\makecell{W/o Memory \\ Initialized}} & \textbf{\makecell{With Memory \\ Initialized}} 
    & \textbf{\makecell{W/o Memory \\ Initialized}} & \textbf{\makecell{With Memory \\ Initialized}} \\
    \midrule
    1  & 3,000 (0)   & 2,458 (626)  & 18.2 (6.0)   & 5.8 (8.0)   & 733 (512)  & 160 (320)  & 0.0\% (0.0\%)    & 55.0\% (51.0\%)  \\
    2  & 2,677 (604) & 2,304 (644)  & 14.5 (13.3)  & 5.3 (8.1)   & 324 (387)  & 255 (349)  & 28.5\% (46.3\%)  & 65.0\% (48.9\%)  \\
    3  & 2,410 (707) & 2,192 (732)  & 10.4 (13.7)  & 3.5 (7.3)   & 257 (457)  & 465 (589)  & 52.4\% (51.1\%)  & 75.0\% (44.4\%)  \\
    4  & 2,475 (699) & 2,150 (672)  & 7.9 (9.5)    & 3.5 (7.0)   & 286 (408)  & 160 (201)  & 47.6\% (51.2\%)  & 75.0\% (44.4\%)  \\
    5  & 2,493 (722) & 2,300 (720)  & 15.9 (12.0)  & 12.0 (9.5)  & 457 (383)  & 470 (508)  & 23.8\% (43.6\%)  & 30.0\% (47.0\%)  \\
    6  & 2,590 (630) & 2,593 (545)  & 19.0 (9.6)   & 14.0 (7.7)  & 476 (648)  & 520 (393)  & 4.7\% (21.8\%)   & 15.0\% (36.6\%)  \\
    7  & 2,503 (555) & 1,994 (784)  & 15.5 (11.9)  & 17.1 (6.9)  & 310 (319)  & 880 (635)  & 23.8\% (43.6\%)  & 5.0\% (22.4\%)   \\
    8  & 2,558 (545) & 2,448 (660)  & 14.5 (11.4)  & 12.5 (6.9)  & 535 (701)  & 1,190 (679) & 25.0\% (44.4\%)  & 10.0\% (30.8\%)  \\
    \midrule
    \textbf{Average} 
    & 2,588 (613)  & \textbf{2,305 (686)}  
    & 14.5 (11.4) & \textbf{9.2 (9.1)}  
    & \textbf{422 (504)}   & 513 (583)  
    & 25.7\% (43.9\%)  & \textbf{41.3\% (49.4\%)}  \\
    \bottomrule
    \end{tabular}
    }
    \caption{Comparison of the robot without initialized memory and the robot with initialized memory. Values are presented as \textit{Average (std)}. Success rate is reported as a percentage (\%), where higher is better; lower is better for the other metrics. Better values for each metric are shown in \textbf{bold}.}
    \label{tab:comparison} 
\end{table*}

Interestingly, the selected CP (Table~\ref{tab:final_cp}) does not reference the ``Brown
rock,'' which only appears in later rounds (Rounds 5–8). Despite this, the robot with
initialized memory continues to facilitate better performance in most of the rounds,
except 7 and 8. At the same time, the decline in the hardest later
rounds is consistent with a mismatch between the reused CP and scenarios in which the
brown rock changes the local risk structure of debris removal.

In a real USAR workflow, such prior memories would likely come from earlier training
exercises, debriefs, or previously documented team routines rather than being learned
from scratch during deployment. Because our memories remain explicit situation-action
structures, they could in principle be reviewed, revised, or disabled by operators
before use. This does not remove the safety risks of aggressive debris-removal
strategies, but it suggests a plausible deployment model in which reusable prior team
knowledge supports coordination while remaining under human oversight.

\section{Related Work}
\label{sec:related_work}

Our work sits at the intersection of memory-based agents, human-robot teaming,
co-learning, and prior MATRX USAR research. Below, we position the paper relative to
these strands and emphasize where our contribution differs.

Prior work on episodic memory and continual learning has shown that storing and revisiting
past experience can support faster adaptation and incremental improvement~\cite{pmlr-v70-pritzel17a,chaudhry2019efficientlifelonglearningagem}.
Related agent-memory architectures have also separated episodic from semantic memory and,
more recently, represented temporally evolving knowledge in graph form for sequential
decision-making~\cite{https://doi.org/10.48550/arxiv.2204.01611,Kim_Cochez_Francois-Lavet_Neerincx_Vossen_2023,kim2026temporalknowledgegraphmemorypartially}.
These works motivate the broader idea that memory should help future decision-making,
but they do not address the specific problem studied here: how to carry forward a prior
human-robot collaboration pattern into a new team episode. More specifically, Diab et
al. study trust-related knowledge transfer in HRI using knowledge graphs, while
Vinanzi et al. show that episodic memory can bias trust and intention reading,
particularly early in interaction~\cite{DiabDemiris2024,VinanziCangelosiGoerick2021}.
Our work differs in both what is stored and how it is reused. Rather than retaining
generic experiences for trust inference or latent policy updates, we represent prior
human-robot collaboration patterns as explicit knowledge-graph memories containing
situation, action-order, and outcome information, and we select one representative
memory for reuse at test time. The contribution is therefore not a general memory
architecture, but a concrete mechanism for transferring an inspectable prior team
experience into a new collaboration episode in the MATRX USAR setting.

Research on human-robot teaming has emphasized mutual adaptation, shared understanding,
team dynamics, trust, and broader human-autonomy teaming requirements~\cite{Nikolaidis,Haripriyan2024TeamDynamics,ChaconQuesada2024SharedMentalModels,Smith2024HumanAutonomyTeaming}.
Related co-learning work has explored user feedback, models of human decision making,
reinforcement learning, and transparency in interactive AI systems~\cite{Kumar2024CoLearning,DBLP:journals/corr/abs-2003-01156,WenskovitchInteractiveAD}.
These studies explain why robots should adapt to human partners, but they typically
focus on adaptation during interaction. In contrast, our emphasis is on pre-interaction
transfer: the robot begins with a selected prior collaboration pattern before meeting a
new partner. This distinction matters because our method targets the earliest stage of
collaboration, when shared routines have not yet formed. In that sense, the paper is
less about learning a better online policy during a session and more about improving
how the robot enters the session in the first place.

Within urban search and rescue, prior MATRX-based studies showed that teams can
externalize collaboration patterns through ontology-supported interfaces and chat, and
that these patterns are useful for analyzing human-robot co-adaptation over repeated
rounds~\cite{10.1145/3434074.3446354,vanZoelen2021,vanZoelen2022}. Complementary work
has also examined how rescuers perceive robotic teammates in disaster
scenarios~\cite{Betta2024RescuerPerceptions}. Our work builds directly on this line,
but changes the role of CPs: instead of treating them only as reflective artifacts or
descriptive interaction traces, we turn them into reusable episodic memories. We then
use graph representation learning to cluster these memories and select a representative
CP for reuse, linking prior collaboration analysis to improved performance in later
human-robot teamwork. This is the key distinction from the earlier MATRX studies: the
same ontology-backed CPs are no longer only analyzed after interaction, but are reused
before interaction to shape future collaboration.

\section{Conclusion}
\label{sec:conclusion}

We presented a study in which a USAR robot is initialized with a
single prior collaboration pattern stored as a knowledge-graph episodic memory. Across
20 participants and 160 round-level observations, the memory-initialized robot raised
victim-rescue success from 25.7\% to 41.3\% and reduced average task time by 283
seconds, with the strongest gains appearing early in interaction. These results suggest
that reusable, inspectable prior team knowledge can help a robot enter collaboration
with a more effective starting point than an empty memory state.

The study remains limited by its simulation setting, small sample, heuristic selection
of a single centroid-nearest CP, and lack of random, expert-selected, or multi-CP
preload baselines. Victim harm also increased in some cases when the robot moved heavy
debris. Future work should compare alternative memory-selection strategies, examine
online memory growth and pruning, and evaluate whether reusable episodic team memory
also improves human-centered outcomes such as trust, coordination quality, cognitive
load, and perceived fluency.

\section*{Acknowledgments}
\label{sec:ack}
This research was (partially) funded by the Hybrid Intelligence Center, a 10-year
program funded by the Dutch Ministry of Education, Culture and Science through the
Netherlands Organization for Scientific Research,
\url{https://www.hybrid-intelligence-centre.nl/}.

\bibliographystyle{ieeetr}
\bibliography{references}

\end{document}